\documentclass[12pt]{iopart}
\usepackage{epsfig}
\usepackage{subfigure}
\begin{document}

\title[Absolute Rb absorption]{Absolute absorption on the rubidium $D$ lines: comparison between theory and experiment}

\author{Paul~Siddons, Charles~S~Adams, Chang Ge and Ifan~G~Hughes}

\address{Department of Physics,
Durham University, South Road, Durham, DH1~3LE, UK}
\ead{paul.siddons@durham.ac.uk}
\date{\today}
\begin{abstract}
\noindent We study the  Doppler-broadened absorption of a weak monochromatic probe beam in a thermal rubidium vapour cell on the $D$ lines. A detailed model of the susceptibility is developed which takes into account the absolute linestrengths of the allowed electric dipole transitions and the motion of the atoms parallel to the probe beam.  All transitions from both hyperfine levels of the ground term of both isotopes are incorporated.   The absorption and refractive index as a function of frequency are expressed in terms of the complementary error function.  The absolute absorption profiles are compared with experiment, and are found to be in excellent agreement provided a sufficiently weak probe beam with an intensity under one thousandth of the saturation intensity is used.  The importance of hyperfine pumping for open transitions is discussed in the context of achieving the weak-probe limit.  Theory and experiment show excellent agreement, with an rms error better than 0.2$\%$ for the $D_2$ line at 16.5$^\circ$C.
\end{abstract}

\maketitle
\section{Introduction}
\label{Intro}
In this paper we develop a model that allows us to predict quantitatively the absorptive and dispersive properties of  rubidium vapour probed in the vicinity of the $D$ lines~\footnote{For an alkali metal atom the $D_2$ transition is $n\,^{2}S_{1/2} \rightarrow n\,^{2}P_{3/2}$, where $n$ is the principal quantum number of the valence electron, and the $D_1$ transition is $n\,^{2}S_{1/2} \rightarrow n\,^{2}P_{1/2}$.}, and compare the predictions with an experimental study of the absolute Doppler-broadened absorption spectrum. In alkali-metal atoms the  $D$ lines have a high oscillator strength, and from an experimental perspective rubidium and cesium are ideal species as they have sufficient room temperature vapour pressure to yield large absorption (10-50\%) in cells of modest length (5-7~cm).  In addition, for these atoms the $D$ lines occur at wavelengths where inexpensive and reliable diode laser sources exist. These transitions are frequently studied in atomic physics; the $D_2$ transition  is used extensively in laser-cooling experiments~\cite{nobel97,Adams97}, whereas  non-linear optical processes such as electromagnetically induced transparency~\cite{ArimondoCPT} and chip-scale atomic magnetometers~\cite{Schwindt04} are realized with the $D_1$ transition.  

Controlling the propagation of light through a medium by modifying its absorptive and dispersive properties is a flourishing area of research~\cite{BoydReview,Fleischhauer05,Milonni02}. Having a model which  calculates the absorption and refractive index of a Doppler-broadened medium is useful for predicting the magnitude of relevant pulse propagation properties, a topical example being  the group delay which enables construction of an all-optical delay line~\cite{Camacho07,Akulshin08} or a slow-light Fourier transform interferometer~\cite{Shi07}.  A model which predicts absolute absorption has a number of applications e.g. in analysing EIT spectra~\cite{Badger01,Moh07}.  Absolute absorption spectroscopy can yield the  number density of the sample being studied and has many applications in physics, chemistry, metallurgy and industry \cite{Corney,Demtroder}; when applied to the measurement of different spectral lines the populations of occupied levels are revealed, from which a temperature can be deduced \cite{Eckbreth}.  In addition, Rb and Cs spectroscopy is frequently used for generating a signal used for frequency reference (``locking") of a laser.  Understanding the evolution of the absorption profile aids in the decision of, e.g. which temperature to use to maximise the signal~\cite{Danny}.  The importance of using a weak probe in order to maximise the absorption will be highlighted below.  Many laser-lock schemes have signals which have a non-trivial dependence of signal amplitude on absorption, e.g., polarization spectroscopy~\cite{Polspec} or the dichroic atomic vapour laser lock (DAVLL) \cite{DAVLL}. The choice of which probe power to use is a trade-off between two competing effects: a weak probe beam ensures that  the largest absorption is obtained, whereas  higher probe power gives a better signal-to-noise ratio.

The aim of this work is to provide a detailed model of the absorption and refractive index for the Doppler-broadened Rb $5S \rightarrow 5P$ transition, and to compare the expected spectral dependence of the absorption with experimental data.  The structure of the remainder of the paper is as follows: Section~\ref{Absorption} explains how to calculate  the expected Doppler-broadened spectra for Rb atoms on the $D$ lines; Section~\ref{Experimental} describes the experimental apparatus and details of the methodology used to measure the absorption profile; Section~\ref{Results} presents   and discusses the results,  and finally, in Section~\ref{Conc}, we draw our conclusions.

\section{Calculating the absorption coefficient of a Doppler-broadened medium}
\label{Absorption}
\subsection{The atomic absorption cross-section}
\label{abscoeff}
The absorption of monochromatic light as it propagates along the $z-$direction through a uniform density atomic vapour is given by the Beer-Lambert law:

\begin{equation}
I(z) = I_0 \exp[-\alpha(\nu,T) z],
\label{eq:beer}
\end{equation}	
where $I(z)$ is the intensity of light at position $z$ inside the medium with an absorption coefficient $ \alpha(\nu,T)$, and  $I_0$ is the  beam intensity at the entrance of the medium.   The absorption coefficient is dependent on the frequency, $\nu $, of incident light and the temperature, $T$, of the medium. We assume that the probe beam is sufficiently weak that the absorption coefficient is independent of intensity.  A full discussion of how weak the light has to be for this simplification to be valid is given in section~\ref{hyperpump}.  

The transmission, $\mathcal{T}$, of a beam through a medium of length $L$ is defined as

\begin{equation}
\mathcal{T} = \frac{I(z=L)}{I_0}=\exp(-\alpha L).
\label{eq:transmission}
\end{equation} 
In general, a medium consists of multiple species, each with multiple transitions.  A beam of light will interact with all species according to

\begin{equation}
\mathcal{T} = \exp[-(\Sigma\alpha_{i})L],
\label{eq:totaltransmission}
\end{equation} 
where the total absorption coefficient, $\Sigma\alpha_{i}, $ is the  sum over $\alpha_{i}$, the absorption coefficient for each transition for each species.

The macroscopic absorption coefficient of the medium can be written~\cite{Loudon} in terms of $\sigma$, the microscopic    atomic absorption cross-section and $\mathcal{N}$,  the number density of the atomic gas, $\alpha = \mathcal{N} \sigma$.
There are two reasons why the medium's absorption coefficient is temperature dependent: (i) the atomic  cross-section is influenced by the Doppler width, proportional to the  square root of the temperature; and (ii) the  number density is a strong function of temperature.  Doppler broadening of the spectral lines is dealt with in section~\ref{velocity}, and the temperature dependence of the atomic density in Appendix A.

We label each hyperfine state of the atom with the usual angular momentum quantum numbers $\vert F_g, m_{F_{g}} \rangle$ for the $^2S_{1/2}$ term, and $\vert F_e, m_{F_{e}} \rangle$ for the $^2P_{3/2}$ or $^2P_{1/2}$ term, where the subscript $g  (e) $ denotes the ground (excited) state.  For a multi-level atom, such as rubidium, the calculation of the atomic cross-section is in two parts: first, the relative linestrengths among the different $\vert F_g, m_{F_g} \rangle \rightarrow \vert F_e, m_{F_e} \rangle$ transitions are calculated, then the absolute value is deduced. These calculations are facilitated by initially assuming the atoms are at rest, with  the manifestation  of atomic motion (Doppler broadening)  incorporated later.  We neglect pressure broadening in these calculations.  Gorris-Neveux {\it et al.} measured the Rb-Rb collisional self-broadening to be of the order of $10^{-7}$~Hz cm$^3$~\cite{Gorris}; for the temperature range spanned in this work the pressure broadening is at least four orders of magnitude less than the natural broadening.  For temperatures greater than approximately $ 120 ^\circ $C the self-broadening becomes comparable to the natural width.
\subsection{Transition frequencies}
In order to predict the absorption spectrum the relative spacing of the hyperfine-resolved energy levels for both Rb isotopes are needed. Zero detuning frequency for  $D_2$ ($D_1$)  is set to be the centre of mass frequency of the 
$5s\,^{2}S_{1/2} \rightarrow 5p\,^{2}P_{3/2}$ ($5s\,^{2}S_{1/2} \rightarrow 5p\,^{2}P_{1/2}$) transition in the absence of hyperfine splitting, taking into account the natural abundance of each isotope. The atomic energy level intervals were obtained for $D_2$ from \cite{Arimondo,Gustafsson,Rapol} and for $D_1$ from \cite{Banerjee}.  The  positions of the atomic transitions relative to the centre of mass for $D_2$ (384,230,426.6~MHz) and $D_1$ (377,107,407.299 MHz) are listed in tables 1\subref{freqD2} and 1\subref{freqD1} respectively.

\begin{table}[ht]
\centering
\caption{Transition frequencies for the (a) $D_2$ line, and (b) $D_1$.}
\begin{tabular}{cc}

\subtable[]{\scriptsize
\label{freqD2}
\centering
\begin{tabular}{@{}ccc}
\br
Line & Detuning / MHz &  $F_e$ \\
\mr
$^{87}$Rb & $-$2735.05\lineup\0 & 1 \\
$F_g = 2\rightarrow F_e = 1,2,3$ & $-$2578.11\lineup\0 & 2 \\
& $-$2311.26\lineup\0 & 3 \\
\mr
$^{85}$Rb & $-$1371.29\lineup\0 & 2 \\
$F_g = 3\rightarrow F_e = 2,3,4$ & $-$1307.87\lineup\0 & 3 \\
& $-$1186.91\lineup\0 & 4 \\
\mr
$^{85}$Rb & \lineup\m 1635.454 & 1 \\
$F_g = 2\rightarrow F_e = 1,2,3$ & \lineup\m 1664.714 & 2 \\
& \lineup\m 1728.134 & 3 \\ 
\mr
$^{87}$Rb & \lineup\m 4027.403 & 0 \\
$F_g = 1\rightarrow F_e = 0,1,2 $ & \lineup\m 4099.625 & 1 \\
& \lineup\m 4256.57\lineup\0 & 2 \\
\br
\end{tabular}
}

&

\subtable[]{\scriptsize
\label{freqD1}
\centering
\begin{tabular}{@{}ccc}
\br
Line & Detuning / MHz &  $F_e$ \\
\mr
$^{87}$Rb & $-$3014.644 & 1 \\
$F_g = 2\rightarrow F_e = 1,2$ & $-$2202.381 & 2 \\
\\
\mr
$^{85}$Rb & $-$1497.657 & 2 \\
$F_g = 3\rightarrow F_e = 2,3$ & $-$1135.721 & 3 \\
\\
\mr
$^{85}$Rb & \lineup\m1538.063 & 2 \\
$F_g = 2\rightarrow F_e = 2,3$ & \lineup\m1900.087 & 3 \\
\\
\mr
$^{87}$Rb & \lineup\m3820.046 & 1 \\
$F_g = 1\rightarrow F_e = 1,2 $ & \lineup\m4632.339 & 2 \\
\\
\br
\end{tabular}
}

\end{tabular}
\label{freqD}
\end{table}

\subsection{Relative linestrength factors}
\label{relative}

The strength of the interaction between an atom and near-resonant electromagnetic radiation is characterized by the dipole matrix elements.  The dipole matrix element of the transition between states $\left|F_g,m_{Fg}\right\rangle $ and $\left|F_e,m_{Fe}\right\rangle $ is $\left\langle F_g,m_{Fg}|er_q|F_e,m_{Fe}\right\rangle $.  In order to calculate this matrix element, it is possible to factor out the angular dependence and write the matrix element as a product of Wigner $3-j$ and $6-j$ symbols and a reduced matrix element \cite{Edmonds,Thompson}.  Thus,

\begin{eqnarray}
\fl\langle F_g,m_{Fg}|er_q|F_e,m_{Fe}\rangle & = & (-1)^{2F_e+\mathcal{I}+J_g+J_e+L_g+S+m_{Fg}+1}\langle L_g||e\textrm{\textbf{r}}||L_e\rangle 
\nonumber \\
&&
\times\sqrt{(2F_g+1)(2F_e+1)(2J_g+1)(2J_e+1)(2L_g+1)} 
\nonumber \\
&&
\times
\Biggl ( \begin{array}{ccc}
F_e & 1 & F_g \\ 
m_{Fe} & -q & -m_{Fg}
\end{array}\Biggr )
\Biggl \{ \begin{array}{ccc}
J_g & J_e & 1 \\ 
F_e & F_g & \mathcal{I}
\end{array}\Biggr \}
\Biggl \{ \begin{array}{ccc}
L_g & L_e & 1 \\ 
J_e & J_g & S
\end{array}\Biggr \}.
\label{eq:dipolematrix}
\end{eqnarray}Here $F, \mathcal{I}, J, L, S, $ and $m_F$ are the angular momentum quantum numbers, and $q$ is the integer change in $m_F$ during the transition.  $\mathcal{I}$, the nuclear spin, has the value $\frac{5}{2}$ and $\frac{3}{2}$ for  $^{85}$Rb and  $^{87}$Rb respectively. $S$, the electron spin, has the value $\frac{1}{2}$.   The $3-j$ symbol is the term contained in the large round brackets, and the $6-j$  in curly brackets.  Note that the $3-j$ symbol is non-zero for $m_{Fe} = m_{Fg} +q$, according to the usual definition of $q$.  $\langle L_g||e\textrm{\textbf{r}}||L_e\rangle$ is the reduced matrix element, and can be expressed in terms of the wavelength of the transition, $\lambda$, and the decay rate of the excited state, $\Gamma$.  By calculating the Wigner coefficients and prefactors, equation~(\ref{eq:dipolematrix}) reduces to

\begin{equation}
\left\langle F_g,m_{Fg}|er_q|F_e,m_{Fe}\right\rangle=c_{m_F}\langle L_g||e\textrm{\textbf{r}}||L_e\rangle \equiv c_{m_F}d,
\label{eq:dipolematrix2}
\end{equation}
where $c_{m_F}$ is a coefficient that determines the transition strength of a particular transition, and is dependent on the initial and final states of the transition.

The strength of a transition is proportional to the square of the transition matrix element, thus the transition strength is $c_{m_F}^2d^2$.  Each hyperfine transition is degenerate in $F$ (since we are assuming zero magnetic field).  
The total transition strength of the hyperfine transition $F_g\rightarrow F_e$ is denoted by $C_F^2 = \Sigma c_{m_F}^2$, where $C_F^2$ is the sum of transition strengths $c_{m_F}^2$ of each Zeeman transition in the hyperfine manifold.  These $C_F^2$ coefficients have been calculated for linearly polarised light ($q=0$), and are tabulated in \ref{Transco}.

\subsection{Absolute absorption coefficient}
\label{absolute}
The reduced matrix element, $d$, can be calculated using the expression for the decay rate \cite{Loudon}

\begin{equation}
\Gamma = \frac{\omega_0^3}{3\pi \epsilon_0 \hbar c^3}\frac{2J_g+1}{2J_e+1}|\langle J_g||e\textrm{\textbf{r}}||J_e\rangle |^2.
\label{eq:gamma}
\end{equation}
$\langle J_g||e\textrm{\textbf{r}}||J_e\rangle $ can be written in terms of $\langle L_g||e\textrm{\textbf{r}}||L_e\rangle $ via the relation

\begin{eqnarray}
\langle J_g||e\textrm{\textbf{r}}||J_e\rangle & = & (-1)^{J_e+L_g+S+1}\langle L_g||e\textrm{\textbf{r}}||L_e\rangle \\
&& \times \sqrt{(2J_e+1)(2L_g+1)} \Biggl \{ \begin{array}{ccc}
L_g & L_e & 1 \\ \nonumber
J_e & J_g & S
\end{array}\Biggr \}.
\label{eq:jl}
\end{eqnarray}
The Wigner $6-j$ coefficient and prefactor, both of which are independent of the $F$ and $m_F$ quantum numbers, can be calculated for the $D_2$ line.  Thus

\begin{equation}
\langle J_g=1/2||e\textrm{\textbf{r}}||J_e=3/2\rangle = \sqrt{\frac{2}{3}}\langle L_g=0||e\textrm{\textbf{r}}||L_e=1\rangle.
\label{eq:jtol}
\end{equation} 
Substituting (\ref{eq:jtol}) into (\ref{eq:gamma}) and rearranging,

\begin{equation}
d=\langle L_g=0||e\textrm{\textbf{r}}||L_e=1\rangle=\sqrt{3}\sqrt{\frac{3\epsilon_0 \hbar \Gamma \lambda^3}{8\pi^2}}\label{eq:reduced}.
\end{equation}
For the   $D_1$ line a similar analysis leads to the same result as equation~(\ref{eq:reduced}).  The reduced dipole matrix element for the fine structure splitting should be identical for the $D$ lines.  However, we have used experimentally measured values for the wavelength and decay rates:  $\lambda = 780.241~\rm{nm}$~\cite{Ye} and $\Gamma = 2\pi \times 6.065 ~\rm{MHz}$~\cite{Volz} for $D_2$, and $\lambda = 794.979~\rm{nm}$~\cite{Barwood} and $\Gamma = 2\pi \times 5.746~\rm{MHz}$~\cite{Volz} for $D_1$.  This yields $d =5.177~ea_0 $ for $D_2$, and $d =5.182~ea_0 $ for $D_1$, where $a_0$ is the Bohr radius.

\subsection{Including atomic velocity}
\label{velocity}
The thermal velocity of atoms along the axis of the probe beam is given by the well-known Maxwell-Boltzmann distribution.  It is Gaussian in nature, with a $1/\rm{e}$ width of $u = \sqrt{2k_{\rm B}T/M}$, where $T$ is the temperature, and $k_{\rm B}$ is the Boltzmann constant and $M$ is the atomic mass.  It is this longitudinal motion that leads to Doppler broadening of the absorption spectra.  Let the angular frequency of the laser be $\omega_{\rm L}$, and that of an atomic resonance be $\omega_{0}$.  The angular detuning, $\Delta$, is defined as $\Delta = \omega_{\rm L} - \omega_{0}$.  For an atom moving along the direction of propagation of the probe beam we incorporate the Doppler effect by simply replacing the detuning by $\Delta -kv$, where $k$ is the magnitude of the wavevector of the light, and $v$ is the atomic velocity. We assume that the experiment is conducted in the weak-probe limit, i.e. the laser intensity is sufficiently low that optical pumping processes which redistribute population amongst the hyperfine levels of the ground term do not occur during the transit of an atom across the finite beam width.  The transverse motion of atoms can therefore be neglected.

\subsection{Electric Susceptibility}
\label{chi}

The  susceptibility, $\chi(\Delta) $, encapsulates both the absorptive and dispersive properties of a medium.  For a medium composed of atoms at rest, the susceptibility for the transition $F_g \rightarrow F_e $ is given by

\begin{equation}
\chi_{F_gF_e}(\Delta) = C_F^2d^2\mathcal{N} \frac{1}{\hbar \epsilon_0}f_\Gamma(\Delta),
\label{eq:chi}
\end{equation}
where $C_F^2d^2$ is the transition strength of the hyperfine transition, $\mathcal{N}$ is the number density, and $f_\Gamma(\Delta)$ is a lineshape factor derived from the steady state solution to the optical Bloch equations of a two-level atom, in the absence of Doppler-broadening.   

As the atomic dipoles are not in phase with the driving light field $\chi(\Delta) $ is, in general, a complex function;  the real part characterises dispersion, and the imaginary the absorption.  The susceptibility for atoms with velocity $v$ along the beam propagation direction is given by~\cite{Loudon}

\begin{eqnarray}
f_\Gamma(\Delta - kv) & = & \frac{i}{\frac{\Gamma}{2}} \biggl[1 - i \Bigl(\frac{\Delta - kv}{ \frac{\Gamma}{2}}\Bigr) \biggr]^{-1} \\\nonumber
& = & \frac{-1}{(\frac{\Gamma}{2})^2} ( \Delta - kv)  \biggl[1 +  \Bigl(\frac{\Delta - kv}{ \frac{\Gamma}{2}}\Bigr)^2 \biggr]^{-1} +  \frac{i}{\frac{\Gamma}{2}} \biggl[1+\Bigl(\frac{\Delta - kv}{ \frac{\Gamma}{2}}\Bigr)^2\biggr]^{-1} \\\nonumber
& \equiv & f_\Gamma^R + if_\Gamma^I.
\end{eqnarray}
$f_\Gamma^R$ and $f_\Gamma^I$ denote the real and imaginary parts of $f_\Gamma$ respectively, with  $f_\Gamma^R$ having a characteristic dispersion profile, and $f_\Gamma^I$ being the Lorentzian absorption profile expected for an homogeneously (natural) broadened system.

The magnitude of the  susceptibility depends on the strength of the transition in question, which is simply a prefactor.  Hence we define for convenience $s(\Delta)$, which is directly proportional to $\chi(\Delta) $, but is independent of the specific atomic transition.  By integrating over the atomic velocity distribution, one obtains the Doppler-broadened lineshape

\begin{equation}
s(\Delta) =\int ^{+\infty}_{-\infty} f_\Gamma(\Delta-kv)\times g_u(v) dv,
\label{eq:vintegral}
\end{equation}  
where

\begin{equation}
g_u(v)  =  \frac{1}{\sqrt{\pi u^2}} \exp \Bigl[- \Bigl(\frac{v}{u} \Bigr)^2 \Bigr],
\label{eq:gau}
\end{equation}
is the normalised Gaussian, with $1/\rm{e}$ width $u$. 
 
Making the substitutions $y=\Delta /ku$, $x=v/u$ and $a=\Gamma/ku$, equation~(\ref{eq:vintegral}) becomes

\begin{equation}
s(y) =\int ^{+\infty}_{-\infty} f_a(y-x)\times g(x) dx,
\end{equation}  
where $s(y)$ is in units of $(ku)^{-1}$, and  $g(x)$ has been normalised, with a dimensionless width of 1. This is in the form of a convolution integral, and can be re-written as
$s(y)=f_a(x)\otimes g(x)$.  Separating the real and imaginary parts of $f_a$,

\begin{equation}
s(y)=f_a^R(x)\otimes g(x) + if_a^I(x)\otimes g(x),
\end{equation}
and hence

\begin{eqnarray}
s^R(y) & = & \textrm{Re}[f_a^R(x)\otimes g(x) + if_a^I(x)\otimes g(x)]\\\nonumber
& = & f_a^R(x)\otimes g(x)\\
\label{eq:S}
s^I(y) & = & \textrm{Im}[f_a^R(x)\otimes g(x) + if_a^I(x)\otimes g(x)]\\\nonumber
& = & f_a^I(x)\otimes g(x),
\end{eqnarray}
where the fact that the convolution of two real functions is real has been used. The imaginary part $s^I$ is related to the absorption coefficient, and is the well-known Voigt function, being the convolution of a Lorentzian and a Gaussian function.

Using the convolution theorem of Fourier transforms, the convolution of two functions can be re-written as

\begin{equation}
S(\tilde y)=F_a(\tilde x)\times G(\tilde x),
\end{equation}
where capitals denote the Fourier transform of a function, and the tilde denotes the reciprocal variable.  The advantage of using this method is that the Fourier transform $S(\tilde y)$ is simply the product of two functions which can be calculated analytically.  $s(y)$ can then be produced from $S(\tilde y)$ by taking its inverse Fourier transform.

For the case of the Voigt profile, the Fourier transforms of $f_a^I(x)$ and $g(x)$ are

\begin{equation}
F_a^I(\tilde x)=\pi \exp \Bigl[- \frac{a}{2} |\tilde x |  \Bigr], 
\end{equation}

\begin{equation}
G(\tilde x)= \exp \Bigl[- \Bigl(\frac{\tilde x}{2} \Bigr)^2 \Bigr].
\end{equation} 
Taking the inverse Fourier transform of their product results in a Voigt profile of

\begin{equation}
s^I(y) =\frac{\sqrt{\pi}}{2} e^{\frac{1}{4} (a-i2y)^2}\biggl(\textrm{Erfc}\Bigl[\frac{a}{2}-iy \Bigr]+e^{i2ay}\textrm{Erfc}\Bigl[\frac{a}{2}+iy \Bigr]\biggr),
\label{eq:erfc}
\end{equation}   
where $s^I(y)$ is in units of $(ku)^{-1}$, and $\textrm{Erfc}[z]$ denotes the complementary error function of $z$.

The real part of the susceptibility is related to the refractive index, and can also be expressed in terms of the complementary error function:
\begin{equation}
s^R(y)=i\frac{\sqrt{\pi}}{2} \biggl( e^{\frac{1}{4} (a-i2y)^2} \textrm{Erfc}\Bigl[\frac{a}{2}-iy \Bigr]-e^{\frac{1}{4} (a+i2y)^2}\textrm{Erfc}\Bigl[\frac{a}{2}+iy \Bigr]\biggr).
\end{equation}
$s^R(y)$ is in units of $(ku)^{-1}$ and, despite the prefactor of $i$, is entirely real.  $s^R(y)$ can  be differentiated with respect to \textit{y} to arbitrary powers to evaluate, e.g., the group refractive index.

\subsection{Absorption Coefficients}
\label{Absorption Coefficients}

The absorption coefficient can be obtained from the imaginary part of the   susceptibility, $\chi(\Delta) $, via

\begin{equation}
\alpha(\Delta) = k\,\rm{Im}[\chi(\Delta)],
\label{eq:alpha}
\end{equation}
where $k$ is the wave number of the probe beam.  $\rm{Im}[\chi(\Delta)] $ has the form of a Voigt profile, $s^I(\Delta)$,  multiplied by prefactors which depend on the properties of the resonant transition. 

The width of the Voigt profile is characterised by a single parameter, $a$: the ratio of the widths of the Lorentzian to the Gaussian profiles. The width of the Lorentzian, $\Gamma $, is the full-width at half-maximum (FWHM) of the hyperfine-free atomic transition.  $\Gamma $ is identical for all hyperfine transitions and Zeeman sublevels within the hyperfine-free manifold, and is also equal for different isotopes of an element.  The width of the Gaussian profile is proportional to the width of the Maxwell-Boltzmann distribution, $u$, and is a function of temperature and isotopic mass.  Considering all of the above, the width of the absorption profile of every hyperfine transition for a particular isotope is identical.

The height of the Voigt profile depends on two factors: the forms of the Lorentzian and Gaussian functions, which are identical for all transitions for a given isotope; and the transition strength of a particular transition.  

All transitions for a particular isotope can be represented by a single Voigt profile, which is then centred on the relevant transition frequencies, and multiplied by the relevant transition strengths.  Hence, recalling equations~(\ref{eq:chi}) and (\ref{eq:alpha}), the absorption profile for a particular hyperfine transition $F_g \rightarrow F_e$ is

\begin{equation}
\alpha_{F_gF_e}(\Delta) = k\,\textrm{Im}[\chi(\Delta)] = k C_F^2d^2\mathcal{N}\frac{1}{2(2\mathcal{I}+1)}\frac{1}{\hbar \epsilon_0}\frac{s^I(y)}{ku}.
\label{eq:alpha2}
\end{equation}
Here, $2\,(2\mathcal{I}+1)$ is the degeneracy of the ground state of the particular isotope (12 for $^{85}$Rb, 8 for $^{87}$Rb).  The degeneracy appears as we are assuming that the population is evenly distributed amongst the ground state Zeeman sublevels (at room temperature the Boltzmann factor is 1 for the two different $F_g$ hyperfine states, and reduces the population of excited states to a negligible level).
The expected transmission profile for the vapour cell can then be calculated as a function of detuning.

\begin{figure}[tbp]
\centering
\includegraphics[width=13cm]{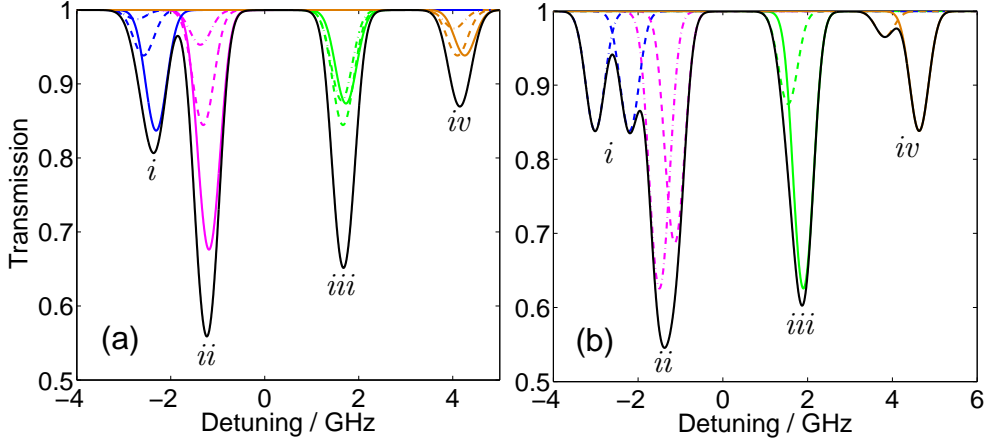}
\caption{Plots of the transmission through a vapour cell of length 75~mm as a function of linear detuning, $\Delta/2\pi$.  Plot (a) shows $D_2$ at $ 20^\circ $C, and (b) shows $D_1$ at $ 30 ^\circ $C.  The blue lines (\textit{i}) show the transmission for the transitions $^{87}$Rb $F_g = 2 \rightarrow F_e $, the magenta (\textit{ii}) $^{85}$Rb $F_g = 3 \rightarrow F_e $, the green (\textit{iii}) $^{85}$Rb $F_g = 2 \rightarrow F_e $,  and the orange (\textit{iv}) $^{87}$Rb $F_g = 1 \rightarrow F_e $.  The solid lines show the transitions between hyperfine states $F_g \rightarrow F_e =F_g +1$, dashed $F_g \rightarrow F_e =F_g $, and dot-dash $F_g \rightarrow F_e =F_g -1$.  The black line shows the total transmission through the cell.  Zero detuning corresponds to the weighted centre of the line.} 
\label{fig1fig}
\end{figure}

Figure~\ref{fig1fig} shows the predicted transmission spectrum for rubidium vapour in a 75~mm-long cell for (a) $D_2$ at $ 20^\circ $C, and (b) $D_1$ at $ 30^\circ $C.  The contributions of the individual $F_g \rightarrow F_e$ transitions are shown, in addition to their combined total.  For both $D$ lines the ground state hyperfine splitting is larger than the Doppler width of $\sim 0.5$~GHz.  For the  $D_2$ line, the excited hyperfine splitting of both isotopes is smaller than the Doppler width; consequently four composite lines are observed. For the $D_1$ line, the excited state splitting for $^{85}$Rb is  smaller than the Doppler width, whereas the splitting for $^{87}$Rb is larger; hence six composite lines are seen.  

\section{Experimental methods and results}
\label{Experimental}
\subsection{Experimental apparatus}

\begin{figure}[tbp]
\centering
\includegraphics[width=12cm]{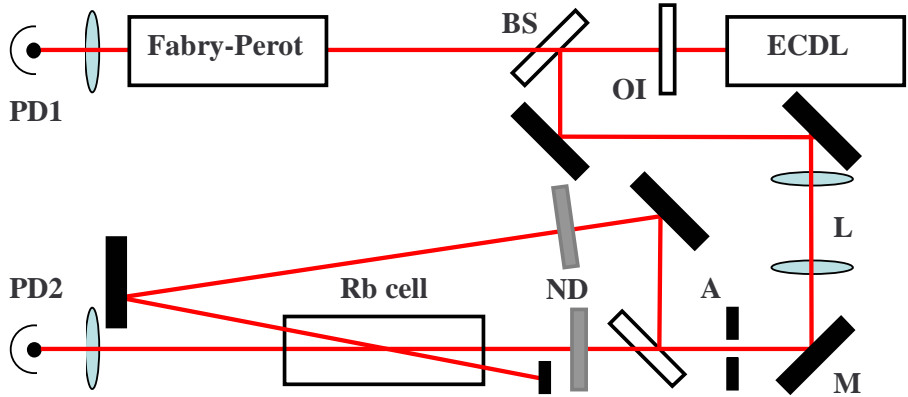}
\caption{Schematic of the experimental apparatus.  Light from an external cavity diode laser (ECDL) passes through an optical isolator (OI) and impinges on a beam splitter (BS).  A fraction of the beam passes through a Fabry-Perot etalon onto a photo detector (PD).  Another fraction of the beam is expanded in a telescope and passes through aperture A.  Mirrors (M) steer the beam and another beam splitter is used to make pump and probe beams which  cross at a small angle in a Rb vapour cell.  The probe beam is incident on a photo detector, and neutral density filters (ND) are used to give independent control over the pump and probe beam powers}
\label{setup}
\end{figure}

We now test the accuracy of the prediction experimentally.  A schematic of the experiment is shown in figure~\ref{setup}.   External cavity diode lasers  were the source of light (Toptica DL100 at 780.2~nm and 795.0~nm for $D_2$ and $D_1$ respectively).
A fraction of the output beam was used as a probe beam for rubidium vapour in a 7.5~cm cell.  A portion of the light was also sent into a Fabry-Perot etalon.  A telescope was used to expand the probe beam before the cell. Before the cell the beam had a radius of (2.00$\pm$0.05)~mm. The cell could be heated to change the vapour pressure of rubidium and hence the opacity.  A thermocouple was used to measure the approximate temperature of the cell.  No attempt was made to null the laboratory magnetic field.  A pump beam generated sub-Doppler spectral features to provide a frequency reference. The crossing angle between
probe and counterpropagating pump within the vapour cell was 6~mrad.  Neutral density filters were used to give independent control of the pump and probe powers.  The Fabry-Perot etalon was used to assist with calibrating and linearizing the frequency scan.  A plane-plane cavity was used, with a separation of the mirrors of 25~cm, with a free-spectral range of 0.60~GHz.  The probe beam was incident on a photo detector comprising a simple current-to-voltage circuit designed to output a voltage linearly proportional to the incident power.

\subsection{Scaling the frequency axis.}
For  the $D_2$ line the frequency axis of the laser scans were linearized by use of the etalon transmission peaks.  In order to generate atomic frequency markers on a scale narrower than the Doppler-broadened features pump-probe spectroscopy was employed.  By counterpropagating a pump  beam with the weak probe it is possible to generate sub-Doppler saturated-absorption and hyperfine-pumping spectra~\cite{satspec}.  For each isotope ground state $F_g$, three transitions $F_g \rightarrow F_e=F_g, F_g\pm1$ are resolved, and so-called cross over resonances~\cite{satspec} are seen halfway between each resonance.  In this way it is possible to obtain 24 atomic resonances. 
\begin{figure}[ht]
\centering
\includegraphics[width=12cm]{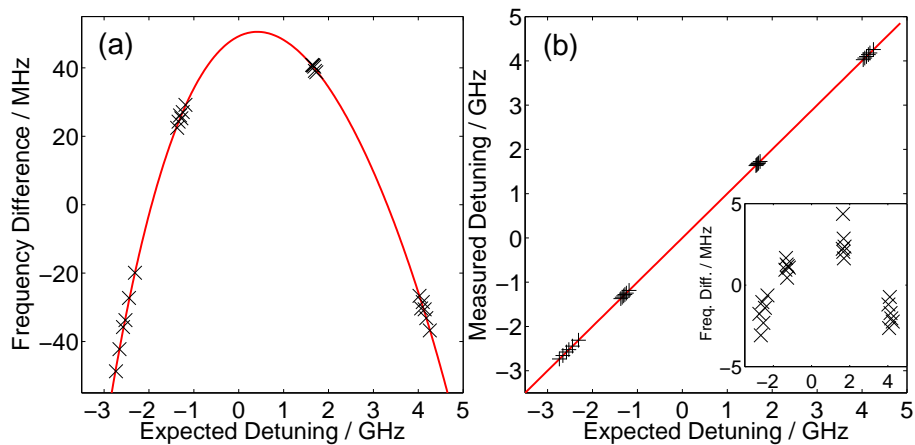}
\caption{(a) Deviation of the measured spectral line frequencies from their expected positions for the $D_2$ line before linearization of the laser scan.  The black crosses mark the measured positions of the sub-Doppler spectra.  (b) Linear fitting of the expected to the measured frequency after linearization.  The red line shows a linear relationship between the two axes, with a gradient of 1 and an intercept of zero.  The inset shows the deviation of the measured spectral line frequencies from their expected positions.}
\label{linear}
\end{figure}

Figure~\ref{linear}(a) shows a plot of the difference between the measured and expected detunings of the 24 atomic resonances before linearization.  The expected detunings were obtained from table~\ref{freqD2}.  The relatively large deviations from zero are seen to have a polynomial relationship with expected frequency.   Figure~\ref{linear}(b) shows a plot of measured detuning of the atomic resonances versus the expected detuning after linearization.  The solid line has slope 1 and passes through the origin.  The inset shows the deviation between measured and expected frequency.  It can be seen that each atomic resonance is within 5~MHz of this ideal fit over a span of 8~GHz.  The residual deviations are a consequence of laser drift.  If a better frequency fitting were desired additional reference etalons could be used.

\begin{figure}[ht]
\centering
\includegraphics[width=12cm]{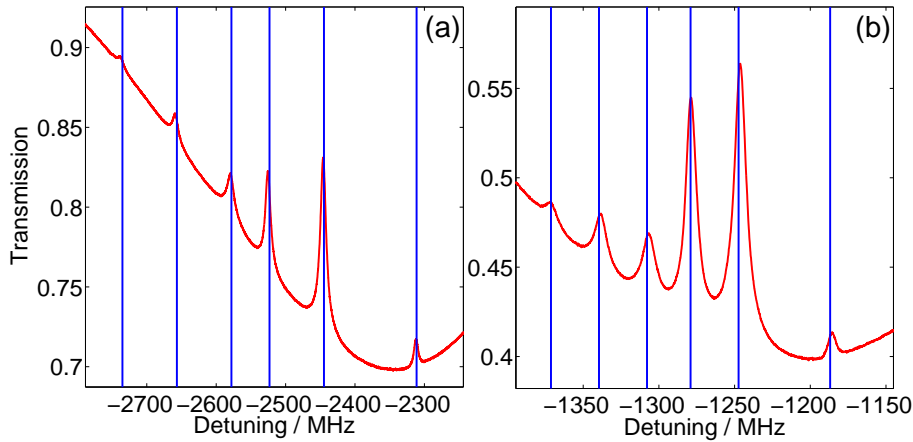}
\caption{Experimentally measured transmission plots for $D_2$ showing saturated-absorption/hyperfine pumping spectra of (a) the $^{87}$Rb $F_g = 2$ line, and (b) the $^{85}$Rb $F_g = 3$.  The vertical reference lines show the expected peak positions.}
\label{subDref}
\end{figure}

Figure~\ref{subDref} shows the pump-probe transmission features for the upper hyperfine level for each isotope; six sub-Doppler features are clearly seen, with their positions being in excellent agreement with the predicted values.

\section{Results}
\label{Results}
\subsection{Effects of hyperfine pumping}
\label{hyperpump}
\begin{figure}[ht]
\centering
\includegraphics[width=12cm]{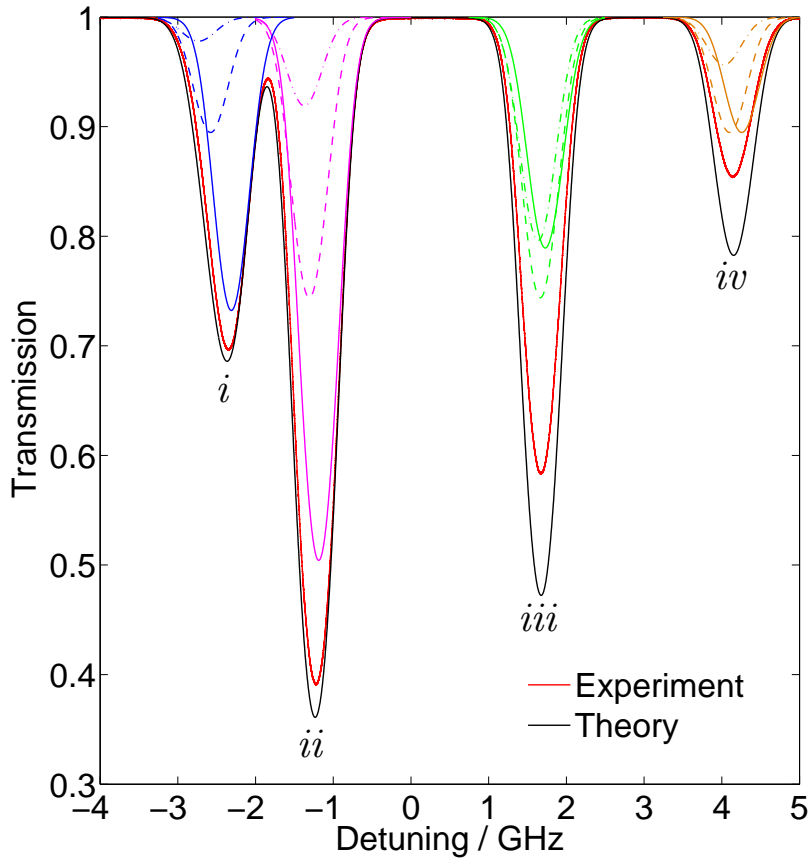}
\caption{Transmission through the vapour cell at $25.4^\circ$C, showing the role of hyperfine pumping in absorption spectroscopy on the $D_2$ line. The solid red line is the experimentally measured transmission and the solid black the predicted value.  Superimposed are the expected transmissions for the individual hyperfine transitions, according to the colour scheme in Figure 1.}
\label{scan}
\end{figure}

Figure~\ref{scan} shows a plot for the $D_2$ line of the transmission, $\mathcal{T}$, versus linear detuning, $\Delta/2\pi$  for a probe intensity of 1.6~$\mu$W/mm$^2$, corresponding to $I/I_{\rm sat}=0.1$.  The expected transmission is also plotted.  The temperature measured using the thermocouple was adjusted at the $0.1^{\circ}$C level in order to fit to the measured data.  Reasonable agreement is obtained for transitions from the upper hyperfine level of the ground term $F_g=\mathcal{I}+1/2 \rightarrow F_e$ (labelled \textit{i} and \textit{ii}), often referred to as the ``laser cooling" transitions.  Poor agreement is seen for transitions from the lower hyperfine level of the ground term $F_g=\mathcal{I}-1/2 \rightarrow F_e$ (\textit{iii} and \textit{iv}), often referred to as the ``repump" transitions.

Although the power of the beam is such that the intensity is less than the saturation intensity, the assumption that the atomic population has not been influenced by the propagation of the probe through the medium is obviously invalid. The process by which an atom is excited from one $F_g$ level and is transferred by spontaneous emission into the other $F_g$ level is known as optical, or hyperfine, pumping.  Allowing for transfer out of the two-level system is known to modify the absorption process~\cite{threelevel}. 

Notice for the Doppler-broadened transitions $^{87}$Rb, \mbox{$F_g = 2 \rightarrow F_e$} (\textit{i}) and $^{85}$Rb, \mbox{$F_g = 3\rightarrow F_e$} (\textit{ii}) in Figure~\ref{scan} the agreement between theory and experiment is excellent on the high-frequency side of the resonance but poor on the low-frequency side - the presence of optical pumping not only reduces the peak absorption but also distorts the lineshape~\cite{Ilkka}.

\begin{figure}[ht]
\centering
\includegraphics[width=15cm]{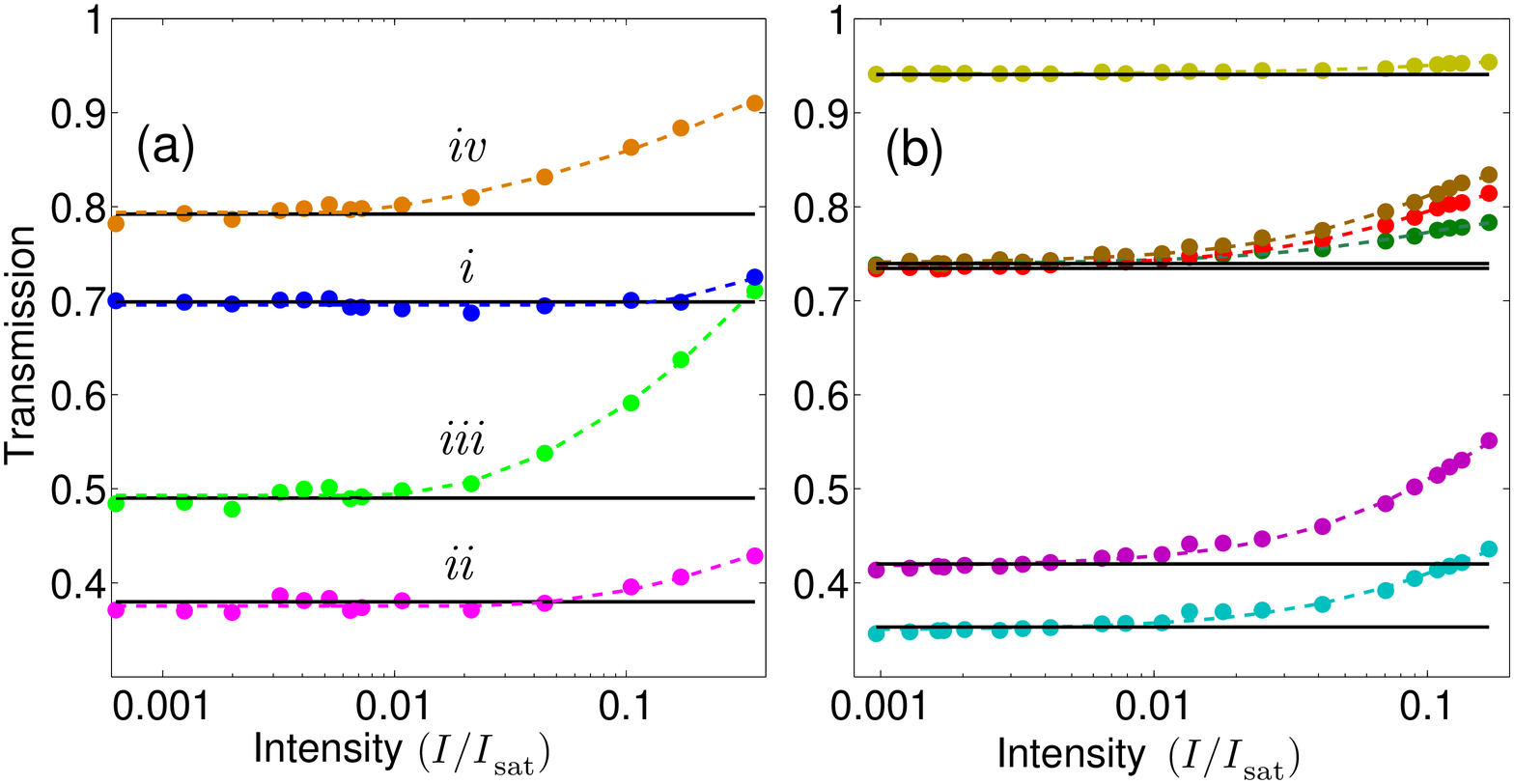}
\caption{(a) The transmission at the centre of the four Doppler-broadened absorption features for $D_2$ is plotted against laser intensity.  The data points correspond to the measured transmission, and the solid lines show the transmission expected.  The dotted lines are guides to the eye.  The cell was at $25^\circ$C.  (b) Similar for $D_1$ where six Doppler-broadened absorption features are observed.  Here the cell was at  $36^\circ$C in order to make the minimum transmission comparable to the $D_2$ line.}
\label{optpump}
\end{figure}

To investigate this further a sequence of spectra were recorded for different probe powers, for both $D$ transitions.  Figure~\ref{optpump} shows the line-centre transmission for (a) the $D_2$ transition in a room temperature cell, and (b) the $D_1$ transition in a cell heated to $36^\circ$C. The laser intensity has been normalized in terms of the saturation intensity~\cite{Corney}.  Consider the closed hyperfine-resolved transition  $D_2$ line: $F_g=\mathcal{I}+1/2 \rightarrow F_e\,=\,\mathcal{I}+3/2$.  Owing to the $\Delta F=0, \pm 1$ selection rule atoms excited into this state have to decay to the ground state from which they started.  These transitions have a significantly larger oscillator strength than the two neighbouring transitions $F_g=\mathcal{I}+1/2 \rightarrow F_e\,=\,\mathcal{I}\pm1/2$, consequently the agreement with  the theory which neglects transfer into other  ground states is good, as was apparent in the high frequency side of the spectrum of the transitions \textit{i} and \textit{ii} in Figure~\ref{scan}.  For the Doppler-broadened transitions \mbox{$^{85}$Rb$\,F_g = 2 \rightarrow F_e$} (\textit{iii}) and  \mbox{$^{87}$Rb$\,F_g = 1 \rightarrow F_e$} (\textit{iv}) there are two closed transitions, $F_g=\mathcal{I}-1/2 \rightarrow F_e\,=\,\mathcal{I}-3/2$.  However these have similar linestrengths to their neighbouring transitions, and hence do not dominate the absorption profile.  There are no such closed transitions in the $D_1$ spectrum.  The conclusion therefore is that great care has to be taken to ensure that the probe beam intensity is sufficiently low that hyperfine pumping does not occur during an atom's transit through the beam - this places a far more strict limitation on the upper intensity to be used in contrast to the condition $I<I_{\rm sat}$ valid for two-level atoms~\cite{Sherlock08}.

\begin{figure}[ht]
\centering
\includegraphics[width=10cm]{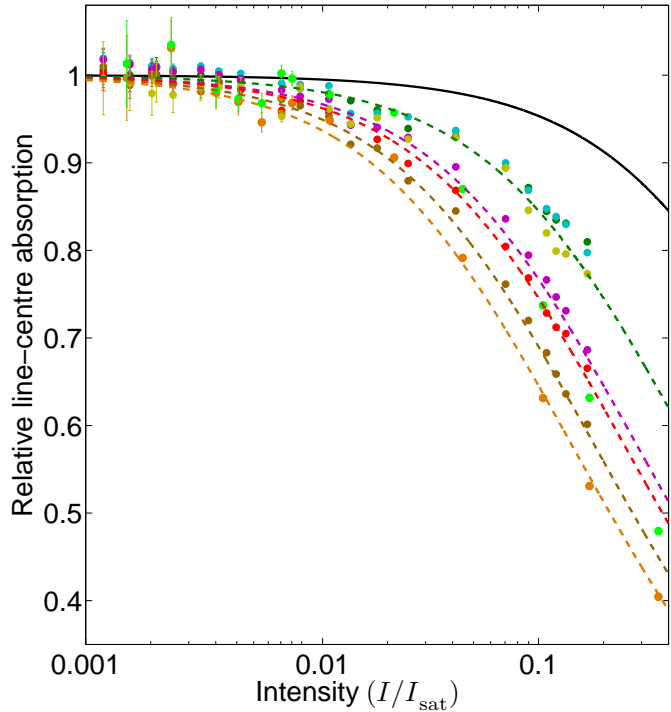}
\caption{Normalized line-centre absorption showing the effects of hyperfine pumping.  The solid line is the theoretical prediction for a Doppler-broadened medium of two-level atoms.  The dotted lines are guides to the eye. Hyperfine pumping on open transitions is seen to be more effective at reducing the line-centre absorption than saturation.}
\label{alphaD}
\end{figure}
An alternative way to visualize the  relative importance of optical pumping is to plot the normalized absorption coefficient $\alpha(I)/\alpha(0)$.  This is done in figure~\ref{alphaD} for the two $D_2$ repump transitions, and all $D_1$ transitions.  Also included is the theoretical prediction for a Doppler-broadened medium consisting of two-level atoms~\cite{Corney} of the form $1/\sqrt{1+I/I_{\rm sat}}$.  It is apparent that optical pumping reduces the absorption at intensities much weaker than those necessary to saturate the transition.  The large error-bars for small intensities are a consequence of the very low light levels and consequently poor signal-to-noise.  The data are fit to curves of the form $1/\sqrt{1+\beta(I/I_{\rm sat})}$, with $\beta$ being a parameter that characterises the effective reduction in  saturation intensity.  This is done as a guide to the eye, and care should be taken not to over interpret this parameterisation.  In this work the beams had a fixed width and the intensity was varied by changing the probe power. It is possible to realise the same intensity with different power  beams of different radii; in this case the presence of optical pumping means that knowledge of intensity alone is not enough to predict the  absorption strength~\cite{Sherlock08}.

\subsection{Comparison of experiment and theory}
\begin{figure}[ht]
\centering
\includegraphics[width=14cm]{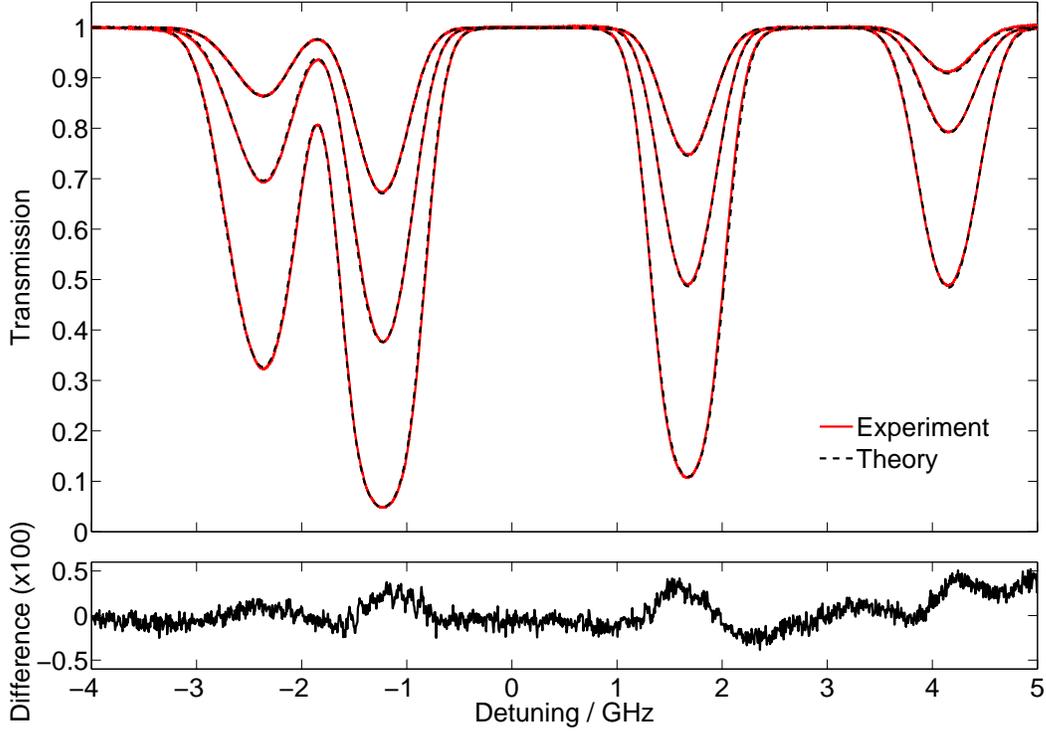}
\caption{Transmission plots for the comparison between experiment and theory, at temperatures of $16.5^\circ$C (top), $25.0^\circ$C (middle), and $36.6^\circ$C (bottom).  Red and black lines show measured and expected transmission respectively.  Below the main figure is a plot of the difference in transmission between theory and experiment for the $16.5^\circ$C measurement. }
\label{mainresult}
\end{figure}

With knowledge of how weak the probe beam had to be, we performed a series of experiments to test the agreement between our theory for the Doppler-broadened absorption profile of rubidium vapour and experiment. The probe intensity was   32~nW/mm$^2$, corresponding to $I/I_{\rm sat}=0.002$.
Figure~\ref{mainresult} shows transmission spectra  at three different temperatures  ($16.5^{\circ}$C, $25.0^{\circ}$C and $36.6^{\circ}$C) for the $D_2$ line. There is excellent agreement between theory and experiment; the rms discrepancy is at the 0.2\% level.  Note that the measured absorption is still slightly smaller than the predicted value.  This could arise due to the broad pedestal of the emission from the laser, and also the finite laser linewidth which is of the order of 0.1\% of the Doppler width.

\section{Conclusion}
\label{Conc}
In summary, we have studied Doppler-broadened spectra for the Rb $D$ lines.  A model was developed which allows the absorption profile and refractive index to be evaluated in terms of tabulated functions.  Excellent agreement was found between theory and experiment for transmissions ranging from 5 to 95\%.   We showed that the effect of hyperfine pumping is significant for open transitions, and outlined how to achieve the weak-field limit.  The weak-probe limit is only reached for $I/I_{\rm sat} \approx 0.001$ for a beam width of 2~mm.  Our model allowing quantitative predictions of the absorption and dispersion in alkali metal vapour will both aid the burgeoning field of controlled light propagation~\cite{Camacho07,Akulshin08,Shi07} and in the understanding of 
the spectra obtained in widely used laser locking schemes~\cite{Danny,Polspec,DAVLL}.

\ack We thank  Aidan Arnold for stimulating discussions.

\appendix

\section{Vapour Pressure and Number Density}
The rubidium vapour cell contains $^{85}$Rb (relative atomic mass $M_{85} = 84.911789738$) and $^{87}$Rb ($M_{87} = 86.909180527$) in their natural abundances of 72.17\% and 27.83\% respectively~\cite{Lide}. 
The vapour pressure (in Torr), $p$, for solid rubidium is given by the following equation \cite{Nese},
\begin{equation}
\fl\textrm{log}_{10}p = -94.04826 - \frac{1961.258}{T} - 0.03771687 \times T + 42.57526 \times \textrm{log}_{10}T,
\label{eq:vpressuresol}
\end{equation}	
and for liquid rubidium is given by
\begin{equation}
\fl\textrm{log}_{10}p =  15.88253 - \frac{4529.635}{T} +0.00058663 \times T -2.99138 \times \textrm{log}_{10}T .
\label{eq:vpressureliq}
\end{equation}	
Using this vapour pressure, the number density, $\mathcal{N}$, of rubidium atoms can be calculated,
\begin{equation}
\mathcal{N} = \frac{133.323 \times p}{k_{\rm B}T}.
\label{eq:ndensity}
\end{equation}
The melting point of rubidium is 39.31 $^{\circ}$C.  The factor of 133.323 converts the vapour pressure from Torr to Pa.  Since there are two isotopes present in the cell, the number densities need to be calculated separately according to their abundance.  

\section{Transition Coefficients}
\label{Transco}

The values of the transition strength factors $C_F^2$ of the $D$ lines are tabulated in this Appendix.

\begin{table}[htb]
\caption{$C_F^2$ for the $D_2$ line of (a) $^{85}$Rb, and (b) $^{87}$Rb.}
\centering
\begin{tabular}{cc}

\subtable[]{
\label{strength85D2}
\centering
\begin{tabular}{@{}ccccc}
\br
 $F_g$ & & $F_e$ &  &   \\

 & 1 & 2 & 3 & 4 \\
\mr
 2 & $\frac{1}{3}$ & $\frac{35}{81}$ & $\frac{28}{81}$ & 0 \\
 3 & 0 & $\frac{10}{81}$ & $\frac{35}{81}$ & 1\\
\br 
\end{tabular}}

&

\subtable[]{
\label{strength87D2}
\centering
\begin{tabular}{@{}ccccc}
\br
 $F_g$ & & $F_e$ &  &   \\
 & 0 & 1 & 2 & 3 \\
\mr
 1 & $\frac{1}{9}$ & $\frac{5}{18}$ & $\frac{5}{18}$ & 0 \\
 2 & 0 & $\frac{1}{18}$ & $\frac{5}{18}$ & $\frac{7}{9}$ \\
\br 
\end{tabular}}
\end{tabular}
\label{strengthD2}
\end{table}

\begin{table}[htb]
\caption{$C_F^2$ for the $D_1$ line of (a) $^{85}$Rb, and (b) $^{87}$Rb.}
\centering
\begin{tabular}{cc}

\subtable[]{
\label{strength85D1}
\centering\begin{tabular}{@{}ccc}
\br
 $F_g$ & $F_e$ &    \\
  & 2 & 3\\
\mr
 2 & $\frac{10}{81}$ & $\frac{35}{81}$ \\
 3 & $\frac{35}{81}$ & $\frac{28}{81}$ \\
\br 
\end{tabular}
}

&

\subtable[]{
\label{strength87D1}
\centering\begin{tabular}{@{}ccc}
\br
 $F_g$ & $F_e$ &   \\
 & 1 & 2 \\
\mr
 1 & $\frac{1}{18}$ & $\frac{5}{18}$ \\
 2 & $\frac{5}{18}$ & $\frac{5}{18}$\\
\br 
\end{tabular}
}
\end{tabular}
\label{strengthD1}
\end{table}

\end{document}